%% file: 0-main.tex
\documentclass[sigconf,nonacm]{acmart}

\usepackage{graphicx}
\usepackage{amsmath}
\usepackage{enumitem}

\newcommand{\name}{Ekar}
\newcommand{\xhdr}[1]{{\noindent\bfseries #1}.}
\DeclareMathOperator{\E}{\mathbb{E}}
\usepackage[flushleft]{threeparttable}
\usepackage{multirow}

\AtBeginDocument{%
  \providecommand\BibTeX{{%
    \normalfont B\kern-0.5em{\scshape i\kern-0.25em b}\kern-0.8em\TeX}}}

\begin{document}

\title{\name: An Explainable Method for Knowledge Aware Recommendation}

\author{Weiping Song}
\authornote{Equal contribution. This work was done when Weiping was visiting Mila.}
\affiliation{\institution{Department of Computer Science, School of EECS, Peking University}}
\email{weiping.song@pku.edu.cn}

\author{Zhijian Duan}
\authornotemark[1]
\affiliation{\institution{Department of Computer Science, School of EECS, Peking University}}
\email{zjduan@pku.edu.cn}

\author{Ziqing Yang}
\affiliation{\institution{Department of Computer Science, School of EECS, Peking University}}
\email{yangziqing@pku.edu.cn}

\author{Hao Zhu}
\affiliation{\institution{Department of Computer Science, School of EECS, Peking University}}
\email{hzhu1998@pku.edu.cn}

\author{Ming Zhang}
\affiliation{\institution{Department of Computer Science, School of EECS, Peking University}}
\email{mzhang_cs@pku.edu.cn}

\author{Jian Tang}
\affiliation{\institution{Mila-Quebec AI Institute, \\HEC Montreal \& CIFAR AI Chair}}
\email{jian.tang@hec.ca}

\begin{abstract}
This paper studies recommender systems with knowledge graphs, which can effectively address the problems of data sparsity and cold start. Recently, a variety of methods have been developed for this problem, which generally try to learn effective representations of users and items and then match items to users according to their representations. Though these methods have been shown quite effective, they lack good explanations, which are critical to recommender systems. In this paper, we take a different route and propose generating recommendations by finding meaningful paths from users to items. Specifically, we formulate the problem as a sequential decision process, where the target user is defined as the initial state, and the edges on the graphs are defined as actions. We shape the rewards according to existing state-of-the-art methods and then train a policy function with policy gradient methods. Experimental results on three real-world datasets show that our proposed method not only provides effective recommendations but also offers good explanations.
\end{abstract}




\maketitle

\input{1-introduction.tex}
\input{2-related-work.tex}
\input{3-definition.tex}
\input{4-model-new.tex}

\input{4-model-plus.tex}

\input{5-experiment.tex}

\input{6-conclusion.tex}

\bibliographystyle{ACM-Reference-Format}
\bibliography{reference}

\end{document}

%% file: 1-introduction.tex
\section{Introduction}

\begin{figure}
{\includegraphics[width=\linewidth]{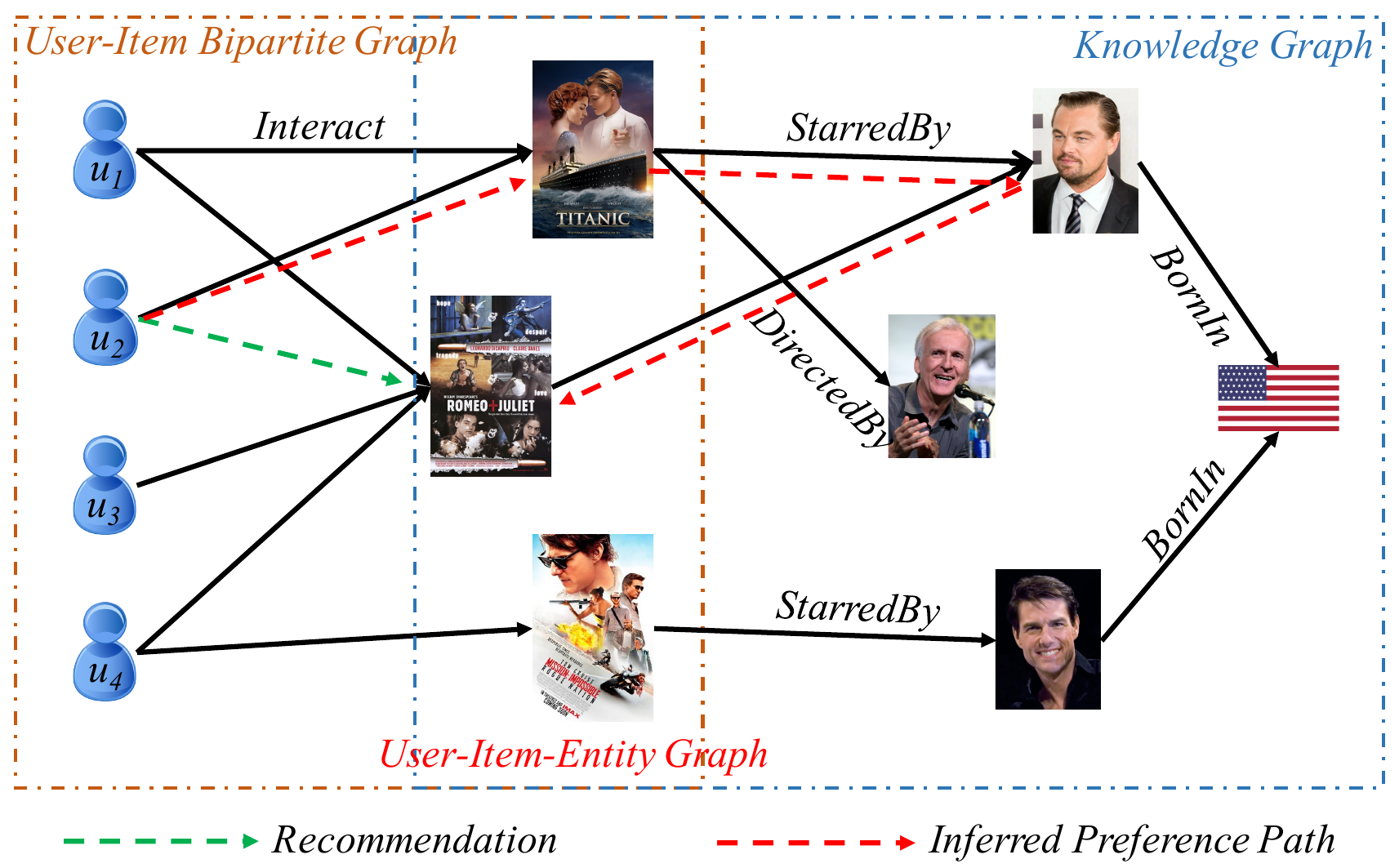}}
{\caption{Explainable recommendation via reasoning over the integrated user-item-entity graph. Our \name\ generates the recommendation (i.e., ``Romeo and Juliet'') by inferring a preference path ``$\text{u}_2\xrightarrow{Interact}\text{Titanic}\xrightarrow{StarredBy}\text{LeonardoDiCaprio}\xrightarrow{StarredIn}\text{Romeo and Juliet}$''.}\label{fig:example}}
\end{figure}

Recommender systems are essential to a variety of online applications such as e-Commerce Websites and social media platforms by providing the right items or information to the users. One critical problem of recommender systems is data sparsity, i.e., some items are purchased, rated, or clicked by only a few users or no users at all. Recently, there is an increasing interest in knowledge graph-based recommender systems, 
since knowledge graphs can provide complementary information to alleviate the problem of data sparsity and have been proved quite useful~\cite{Cao_2019,Ma_2019,wang2018ripplenet,wang2018dkn,zhang2016collaborative,zhang2018explainable}. 

Generally, 
existing knowledge graph-based recommendation methods try to learn effective representations of users and items according to the user-item interaction graphs and item-entity knowledge graphs, and then match the items to the users according to learned representations. For example,~\citet{zhang2016collaborative} learn item representations by combining their representations in user-item graphs and knowledge-graphs. 
~\citet{zhang2018learning} learn user and item representations on the integrated user-item-entity graphs based on knowledge graph embedding (KGE) method like TransE~\cite{bordes2013translating}. ~\citet{Wang_2019} and \citet{Cao_2019} jointly optimize the recommendation and knowledge graph embedding tasks in a multi-task learning setting via sharing item representations. These methods have been proved quite effective by integrating information from both user behaviors and knowledge graphs.

Although these methods are very effective, they lack good explanations. 
Intuitively, if the recommender system can give an explanation of a recommendation, the users would have more interest and trust in the recommended item~\cite{herlocker2000explaining, lu2018like,zhang2018explainable,zhang2014explicit}. 
Indeed, there is some existing work that aims to provide such explanations for the recommendation results. For example, the RippleNet~\cite{wang2018ripplenet} aims to explain the recommendations by analyzing the attention scores. However, their method relies on the post analysis of the soft attention scores, which may not always be trustworthy. 

In this paper, we take a different route and propose to generate a path from the target user to relevant items in the integrated user-item-entity graph. Take the movie recommendation in Figure~\ref{fig:example} as an example. For the target user $u_2$, a path (the red dashed lines) is generated to the item "Romeo and Juliet" since 1) user $u_2$ watched movie ``Titanic''; 2) ``Titanic'' is starred by ``Leonardo DiCaprio''; and 3) ``Leonardo DiCaprio'' also stars in ``Romeo and Juliet''. We can see that such a path
offers good explanations of the recommendations in addition to provide meaningful recommendations.

However, finding meaningful paths on the large user-item-entity graph is challenging. One may enumerate all paths between user-item pairs and then use a classification/ranking model to select the most meaningful paths~\cite{Wang2018ExplainableRO}. Nevertheless, enumerating paths between users and items is intractable due to the exponentially large path space in the user-item-entity graph. Although sampling a certain number of paths via breadth-first-search can be a practical substitution to enumerating, it has no assurance on the meaningfulness of sampled paths. In this paper, we instead formulate the generation of meaningful user-to-item paths as a sequential decision process. Specifically, the recommender agent starts from target users and extends its paths to relevant items by sequentially selecting walks on the user-item-entity graph. During training, we assign each path a positive reward if the starting user and terminal entity constitute an observation in recommendation data. Considering that the reward could be extremely sparse at the beginning due to the huge exploration space, we further augment the reward signals by reward shaping~\cite{ng1999policy}, where a soft reward function is first learned using state-of-the-art knowledge graph embedding methods~\cite{yang2014embedding,dettmers2018convolutional,sun2019rotate}. We use the REINFORCE~\cite{williams1992simple} algorithm to maximize the expected rewards of our recommender agent. Finally, we verify the effectiveness and the explainability of the proposed method on three real-world datasets. Quantitative results demonstrate that our proposed \name \footnote{\name\ represents \textbf{E}xplainable \textbf{k}nowledge \textbf{a}ware \textbf{r}ecommendation.} model: 1) significantly outperforms existing state-of-the-art KG-based recommendation methods, 2) offers clear and convincing explanations in the form of meaningful paths from users to recommended items.

To summarize, the contributions of this work are as follows:

\begin{itemize}[leftmargin=*]
    \item We propose generating (rather than discriminating) meaningful paths from users to items, where the end of paths (i.e., items) are recommendations and the paths themselves are explanations w.r.t. the recommendations.
    \item We propose a novel deep reinforcement learning-based approach to infer such meaningful paths. To encourage exploration and stabilize training, we design a well-shaped reward function based on existing knowledge graph representation learning methods.
    \item We conduct extensive experiments to validate the proposed approach. Experimental results show that our model significantly outperforms previous state-of-the-art method (KTUP~\cite{Cao_2019}) with at most 31.31\% hit ratio gain for top-N recommendation. What's more, studies on recommendation paths show good explainability of our approach.
\end{itemize}

\noindent\textbf{Organization.} Section 2 reviews related work. In Section 3, we give formal definitions used in this paper. Section 4 introduces our proposed deep reinforcement learning-based approach for explainable knowledge graph-based recommendation. We introduce two mechanisms for action selection in Section 5. Extensive experimental results and analysis are presented in Section~\ref{sec::experiment}, followed by the conclusion of our work in Section~\ref{sec::conclusion}.

%% file: 2-related-work.tex
\section{Related work}
Our work is conceptually related to the explainable recommendation, knowledge graph-based recommendation, and recent advancements in applying reinforcement learning into relational reasoning.

\subsection{Explainable Recommendation} 
As a widespread concern in the AI community, explainability has been widely discussed in recommender systems.
According to Zhang and Chen~\cite{zhang2018explainable}, most of the existing explainable recommendation methods typically provide explanations via identifying users' preference on item features~\cite{bauman2017aspect,wang2018explainable,zhang2014explicit}, understanding latent factors with topic modeling~\cite{mcauley2013hidden, wu2015flame,zhao2015sar}, or ranking over the user-item-aspect graph~\cite{he2015trirank}. However, these methods require external information (e.g., reviews) about items, which may be difficult to collect. Some recent advancements utilize the
attention mechanism~\cite{li2017neural,song2019session} to provide explanations, but they need extra efforts to explore attention scores.
Since knowledge graphs (KG) provide common knowledge about our world, many recent works use knowledge graphs~\cite{Cao_2019,wang2018ripplenet,wang2018dkn,zhang2016collaborative} to provide explainable recommendations, which will be further discussed in the following paragraph.

\subsection{KG-based Recommendation} Our work is closely related to KG-based recommendation, which utilizes general knowledge graphs (e.g., DBpedia, YAGO, and Satori) to improve recommender systems. Existing KG-based recommendation methods can be roughly divided into two classes: embedding-based methods and path-based methods. In embedding-based methods, users and items are represented by low-dimensional vectors, where entities' embeddings from the knowledge graph are used to enhance corresponding items' representations~\cite{Cao_2019,wang2018dkn,Wang_2019,zhang2016collaborative,10.1145/3292500.3330989}. Although these methods perform well, it's hard to explain the recommendation results because representations are in a latent space. In path-based methods, meta-paths and meta-graphs are commonly used to extract various semantic dependencies between users and items~\cite{yu2014personalized,zhao2017meta}. However, it is almost
computationally infeasible to enumerate all the useful meta-paths or meta-graphs. Moreover, the meta-paths and meta-graphs 
need to be manually defined and cannot generalize to new datasets. Instead of pre-defining specific paths, the RippleNet~\cite{wang2018ripplenet} directly propagates users' preferences along edges in KG via the attention mechanism and then interprets the recommendations according to the attention scores, which however might not be trustworthy. The most recent work KPRN~\cite{Wang2018ExplainableRO} uses LSTM to model the paths between users and items. However, sampling paths via breadth-first-search (BFS) is inefficient and may miss meaningful paths.
Different from KPRN, our method defines the path finding problem as a sequential decision problem. We train an agent to automatically generate a meaningful path between a user and his/her relevant item via policy gradient methods.
A concurrent work to ours is PGPR~\cite{Xian:2019:RKG:3331184.3331203}, which also formulate the recommendation task as a sequential decision process over knowledge graphs. The major differences are two folds: 1) we encode the complete path history as current state while PGPR only considers a small portion of it; 2) we first pre-train a state-of-the-art KGE model for reward shaping and therefore achieve much better performance.

\subsection{Relational Reasoning with Reinforcement Learning} Our work is also related to recent work on knowledge graph reasoning with reinforcement learning~\cite{das2017go,lin2018multi,xiong2017deeppath}, which aims to train an agent to walk on knowledge graphs to predict the missing facts. However,
their goal is different from ours. We focus on the problem of recommendation with knowledge graphs and aim at finding meaningful paths for explaining the recommendation results while they focus on facts prediction.

%% file: 3-definition.tex
\section{Definitions}
To relieve the data sparsity issue, incorporating auxiliary knowledge about items has been attracting increasing attention\cite{Cao_2019,wang2018ripplenet,wang2018dkn,zhang2016collaborative,zhang2018explainable}. Typically external knowledge is represented by a graph, where the nodes are entities and the edges are relations between two entities. Since some of entities in knowledge graphs and some of items in user-item graphs can be aligned, we can merge the knowledge graphs and the user-item graphs into an integrated user-item-entity graph, which is formally defined as follows:

\vspace{3pt}

DEFINITION 1. \textbf{(User-item-entity Graph)} Let $\mathcal{G}=(\mathcal{U},\mathcal{I})$ denote the user-item bipartite graph, where $\mathcal{U}$ is the set of users, and $\mathcal{I}$ is the set of items.
Besides, we also have access to an open knowledge graph $\mathcal{G}_k=(\mathcal{E}_k, \mathcal{R}_k)$, where $\mathcal{E}_k$ is the entity set and $\mathcal{R}_k$ is the relation set. Each triplet $<e_h, r, e_t>\in \mathcal{G}_k$ indicates there exists a relation $r\in \mathcal{R}_k$ from head entity $e_h\in \mathcal{E}_k$ to tail entity $e_t\in \mathcal{E}_k$. For example, $<\textit{Titanic, DirectedBy, JamesCameron}>$ reflects the fact that ``Titanic'' is directed by ``James Cameron''. 
As some items/entities are shared in $\mathcal{I}$ and $\mathcal{E}_k$, we merge the user-item bipartite graph $\mathcal{G}$ and the knowledge graph $\mathcal{G}_k$ into an integrated \textit{user-item-entity graph} $\mathcal{G}'=(\mathcal{V}', \mathcal{R}')$, where $\mathcal{V}'=\mathcal{U}\cup\mathcal{I}\cup\mathcal{E}_k$. For the user-item interaction graph $\mathcal{G}$, we assume all the edges belong to a special relation ``Interact'', and therefore $\mathcal{R}'=\{``\mbox{Interact}"\}\cup\mathcal{R}_k$. An example of such user-item-entity graph could be Figure~\ref{fig:example}.

Typically users do not consume items at random. In other words, there should be \textit{implicit reasons} why a specific item is selected by a user. As shown by the red line in Figure~\ref{fig:example}, a user may choose to watch movies starred by the same actor who showed superb acting skills in the movie they watched before. To improve users' acceptance of recommendations, we therefore propose to infer such meaningful user-to-item paths to \textit{explicitly} provide explanations. We define the resulting path-based explainable recommendation problem as follows:

\vspace{3pt}

DEFINITION 2. \textbf{(Path-based explainable recommendation}) Given a user $u$, \textit{path-based explainable recommendation} task aims to generate a set of paths from $u$ to relevant items on the user-item-entity graph $\mathcal{G}'$. Such paths not only allow to find the relevant items but also offer good explanations.

%% file: 4-model-new.tex
\section{methodology}

\begin{figure*}
\centering
\includegraphics[width=0.96\linewidth]{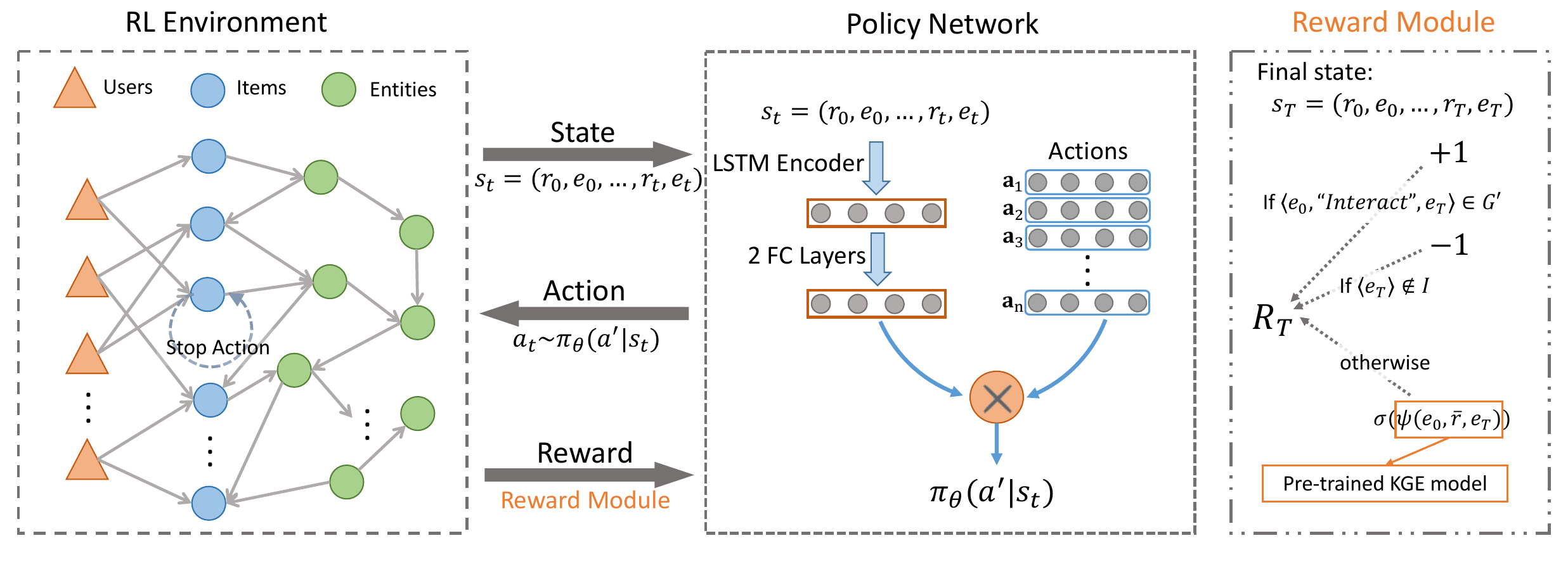}
\vspace{-10pt}
\caption{A schematic view of our proposed \name\ model. We formulate the generation of explainable paths as a sequential decision process. The user-item-entity graph is the RL environment and a policy network interacts with it. To stabilize training and encourage the agent to explore diverse paths, we design a well-shaped reward function.}\label{fig::model}
\end{figure*}

In this section, we describe our proposed \name\ model for path-based explainable recommendation task in detail.

Overall, we formulate the generation of explainable paths as a Markov Decision Process (MDP) on the user-item-entity graph, and we use deep reinforcement learning to solve it. As shown in Figure~\ref{fig::model}, the user-item-entity is treated as the environment, from which we get the observation $s_t$, i.e., a sequence of visited nodes and edges. Based on the encoded state $\mathbf{s}_t$, a policy network consisting of two fully-connected layers outputs the probability distribution over possible action space. Finally, we define the reward being $+1$ if our agent successfully finds those items consumed by the target users $u$ in history. Furthermore, we augment the rewards based on pre-trained state-of-the-art KGE models for the purpose of stabilizing training and encouraging the agent to explore diverse recommendation paths.

\subsection{Formulating Recommendation as a Markov Decision Process}\label{subsec::mdp}
There are some existing methods~\cite{Wang2018ExplainableRO} trying to find meaningful paths between users and items. These methods first sample a collection of paths with breadth-first or depth-first search strategy and then measure the meaningfulness of the paths with a classification or ranking model. However, the number of possible paths between a user and an item could be exponentially large, and sampling a few of them could miss the meaningful ones. 
Moreover, the paths sampled through BFS or DFS strategy may not always be meaningful. 
In this paper, we take a different route and formulate the problem as a sequential decision making problem on the user-item-entity graph $\mathcal{G}'$. We aim to train an agent to walk on $\mathcal{G}'$ to find relevant items.
Starting from a target user $u$, the agent sequentially selects the next neighbor on the integrated user-item-entity graph $\mathcal{G}'$ until it reaches the predefined maximum number of steps $T$.
Formally, we define the states, actions, transition, and rewards of the Markov Decision Process as follows:

\vspace{3pt}
\xhdr{States} We represent the state as the sequence of traversed relations and entities so far, i.e., $s_t=(r_0, e_0, r_1, e_1, ..., r_t, e_t)\in \mathcal{S}_t$, where $r_t\in \mathcal{R}'$ and $e_t\in \mathcal{V}'$ are relations and entities respectively. 
The initial state $s_0=(r_0, e_0)$ represents the target user, and $r_0$ is an artificially introduced relation to be consistent with other $(r_t, e_t)$ pairs.

\vspace{3pt}
\xhdr{Actions} When the agent is under state $s_t$, it can choose an outgoing edge of entity $e_t$ as its next action.
Formally, we define the possible actions under state $s_t$ as $\mathcal{A}_t=\{a=(r', e')|(e_t, r', e')\in \mathcal{G}'\}$.

\vspace{3pt}
\xhdr{Transition} For the state transition $\mathcal{P}(\mathcal{S}_{t+1}=s|\mathcal{S}_t=s_t, \mathcal{A}_t=a_t)$, we adopt a deterministic strategy and simply extend the current state $s_t$ by adding the new action $a_t=(r_{t+1}, e_{t+1})$ as the next state, i.e., $s_{t+1}=(r_0, e_0, ..., r_t, e_t, r_{t+1}, e_{t+1})$.

\vspace{3pt}
\xhdr{Rewards} No intermediate reward is provided for $(s_{t<T}, a_{t<T})$. The final reward depends on whether or not the agent correctly finds interacted items of the user $u$.
Given the terminal entity $e_T$, the final reward $R_T$ is $+1$ if user $e_0$ has interacted with $e_T$, $0$ if $e_T$ is an item but user $e_0$ has not interacted with it and $-1$ if $e_T$ is not an item-type entity.

\subsection{Solving Recommendation MDP with Policy-based Reinforcement Learning}\label{subsec::drl}
Next, we introduce a deep policy-based reinforcement learning method to solve the above MDP. The learned policy is then used to generate recommendations and explanations during inference.

\subsubsection{Architecture of Policy Network} Since there are usually millions of entities and hundreds of relations in user-item-entity graph $\mathcal{G}'$, it is almost impossible to utilize discrete states and actions directly, the number of which is exponential to the number of symbolic atoms in $s_t$ and $a_t$ respectively. We, therefore, choose to represent entities and relations in $\mathcal{G}'$ with low-dimensional embeddings. Each action $a=(r, e)$ is represented as the concatenation of relation and entity embeddings, i.e., $\mathbf{a}=[\mathbf{r}';\mathbf{e}']$. The state $s_t=(r_0, e_0, ..., r_t, e_t)$ is encoded by an LSTM~\cite{hochreiter1997long}:
\begin{equation}
\begin{aligned}   
    \mathbf{s}_{0} &= \textit{LSTM}(\mathbf{0}, [\mathbf{r}_0; \mathbf{e}_0]), \\
    \mathbf{s}_t &= \textit{LSTM}(\mathbf{s}_{t-1}, [\mathbf{r}_{t}; \mathbf{e}_t]), t>0\\
\end{aligned}
\end{equation}
where $\mathbf{0}$ is a zero vector and $\mathbf{s}_t$ is the low-dimensional representation of state $s_t$. 

Based on the parameterized state $\mathbf{s}_t$ and the parameterized action $\mathbf{a}$, we calculate the probability distribution over possible action space $\mathcal{A}_t$ as follows:
\begin{equation}
    \begin{aligned}
    & \mathbf{y}_t = \mathbf{W}_2 \textit{ReLU}(\mathbf{W}_1\mathbf{s}_t + \mathbf{b}_1) + \mathbf{b}_2, \\
    & \pi_\theta(a'|s_t) = \frac{\exp({\mathbf{a}'}^\top \mathbf{y}_t)}{\sum_{a\in \mathcal{A}_t}\exp(\mathbf{a}^\top \mathbf{y}_t)},
    \end{aligned}
\end{equation}
where \{$\mathbf{W}_1, \mathbf{W}_2$\} and \{$\mathbf{b}_1, \mathbf{b}_2$\} are weight matrices and weight vectors of a two-layer fully-connected neural network, $\textit{ReLU}(x)=\textit{max}(0, x)$ is the non-linear activation function and $\pi_\theta(a'|s_t)$ is the probability of taken action $a'$ under state $s_t$.

\subsubsection{Reward Augmentation}
According to our initial definition of rewards, the agent gets a positive reward if and only if it successfully finds the target item. However, this might be problematic for a few reasons: First, for a large user-item-entity graph $\mathcal{G}'$, it is very difficult for the agent to reach the correct items due to the huge search space, especially at the beginning of training~\cite{xiong2017deeppath}. In other words, the rewards will be very sparse. As a result, the learning process of the agent could be very inefficient and take a long time to converge. Second, the goal of recommender systems is to infer new items that users are likely to interact with in the future, rather than repeating users' historical items. However, receiving positive rewards only from historical items discourages the agent to explore new paths and items, which should be the target of recommender systems. To accelerate the training process and meanwhile encourage the agent to explore items that have not been purchased or rated by the target user, 
we propose to shape the rewards~\cite{ng1999policy} in the following way: 
\begin{equation}
    R_T = 
    \begin{cases}
    1, & \text{if } e_T\in \mathcal{I} \text{ and } <e_0, \bar{r}, e_T> \in \mathcal{G}', \\
    \sigma(\psi(e_0, \bar{r}, e_T)),
    & \text{if } e_T\in \mathcal{I} \text{ and } <e_0, \bar{r}, e_T> \notin \mathcal{G}', \\
    -1 , & \text{otherwise},
    \end{cases}
    \label{eq:reward}
\end{equation}
where $\sigma(x)=\frac{1}{1+e^{(-x)}}$ is the sigmoid function and $\bar{r}$ represents the relation ``Interact''.
$\psi(e_0, \bar{r}, e_T)$ is the score function that measures the correlation between user $e_0$ and the searched item $e_T$. In our study, $\psi(e_0, \bar{r}, e_T)$ is pre-trained by maximizing the likelihood of all triplets in graph $\mathcal{G}'$ and can be the score function of any state-of-the-art knowledge graph embedding models~\cite{dettmers2018convolutional,yang2014embedding,sun2019rotate}. For example, the score function is $\psi(e_0, \bar{r}, e_T)=-\|e_0 + \bar{r} - e_T \|$ in TransE~\cite{bordes2013translating} model.
Different from the original rewards defined in Section~\ref{subsec::mdp}, items that the target user has not interacted with now receive positive rewards, which are determined by the pre-trained knowledge graph embeddings.

\subsubsection{Optimization} Finally, we use the policy gradient~\cite{sutton2000policy} method to optimize our policy network. During training, the agent starts with an initial state $(r_0, e_0)$, where $e_0$ is the target user, and sequentially extends its path to a maximum length of $T$. We then use the reward function (i.e., Eq.~\ref{eq:reward}) to assign the trajectory $(s_0, a_0, s_1, a_1, ..., s_T)$ a final reward. Formally, we define the expected rewards over all traversed paths of all users as:
\begin{equation}
    J(\theta)= \E_{e_0\in \mathcal{U}}[\E_{a_1, a_2, ..., a_T\sim \pi_\theta(a_t|s_t)}[R_T]].
\end{equation}
which is maximized via gradient ascent, and the gradients of all parameters $\theta$ are derived by the REINFORCE~\cite{williams1992simple} algorithm, i.e.,
\begin{equation}
    \bigtriangledown_{\theta} J(\theta) \approx \bigtriangledown_{\theta} \sum_{t} R_T\log \pi_\theta(a_t|s_t). \\
\end{equation}

%% file: 4-model-plus.tex
\begin{table*}[th]
\centering
\scalebox{1.0}
{
\begin{tabular}{lccccccc} 
\toprule
\multirow{2}{*}{Data} & \multicolumn{4}{c}{User-Item Interaction} & \multicolumn{3}{c}{Knowledge Graph}\\\cmidrule(r{1em}l{1em}){2-5}\cmidrule(r{1em}l{1em}){6-8}
    & \# Users & \# Items & \# Events & Sparsity & \# Entities & \# Relations & \# Triplets  \\
\midrule
Last.FM & 1,872 & 3,846 & 21,173 & 99.71\% & 9,366 & 60 & 15,518 \\
MovieLens-1M & 6,040 & 2,347 & 656,462 & 95.37\% & 7,008 & 7 & 20,782 \\
DBbook2014 & 5,576 & 2,598 & 65,445 & 99.55\% & 10,149 & 13 & 135,580 \\
\bottomrule
\end{tabular}
}
\caption{Statistics of evaluation datasets and corresponding knowledge graphs.}
\label{tab::dataset}
\end{table*}

\section{Further Constraints on Actions}

The current action space $\mathcal{A}_t$ under state $s_t$ is defined as the set of outgoing edges of current entity $e_t$. This could be problematic for two reasons. First, for $t< T$, if entity $e_t$ is already the correct item (i.e., $(e_0, e_t)\in \mathcal{G}$), the agent should stop and not continue to walk to other entities. Second, since the REINFORCE algorithm tends to encourage the agent to repeat historical experiences which receive high rewards~\cite{Guu_2017}, the algorithm may discourage the agent from exploring new paths and items, which could be relevant to the target user. We address the two problems in the following ways:


\subsection{Stop Action} As the length of paths may vary for different user-item pairs, we should provide the agent an option to automatically terminate when it believes that it has found the right items ahead of $T$.
Following~\citet{das2017go}and~\citet{lin2018multi}, we add a special link from each node to itself. In this way, we allow the agent to stay at the ground truths, which can be understood as a stop action. We show the impact of using stop action in Section~\ref{subsec:ablation} by setting different path length $T$.

\subsection{Action Dropout} To prevent the agent from repeating historical high-reward paths and encourage it to explore more possibilities, we propose to use action dropout~\cite{lin2018multi} during the training. Specifically, instead of sampling an action from original $\pi_\theta(a_t|s_t)$, we use a mask upon $\pi_\theta(a_t|s_t)$ to randomly drop some actions. In addition, action dropout can also help alleviate the problem of irrelevant paths between a user and an item since these paths may be found coincidentally at the beginning of training.


%% file: 5-experiment.tex
\section{Experiments}\label{sec::experiment}

\begin{table*}[t!]

\scalebox{1.0}
{
    \begin{tabular}{lcccccc}
    \toprule
    \multirow{2}{*}{Model} & \multicolumn{2}{c}{Last.FM} & \multicolumn{2}{c}{MovieLens-1M} &  \multicolumn{2}{c}{DBbook2014} \\
    & HR@10 & NDCG@10 & HR@10 & NDCG@10 & HR@10 & NDCG@10 \\
    \midrule
    ItemKNN & 0.0605 & 0.0511 & 0.0738 & 0.2273 & 0.0702 & 0.0665 \\
    BPR-MF & 0.1199 & 0.0916 & 0.0895 & 0.1914 & 0.0829 & 0.0565 \\
    \midrule
    RippleNet & 0.1008 & 0.0641 & 0.1269 & 0.2516 & 0.0763 & 0.0571 \\
    CFKG & 0.1781 & 0.1226 & 0.1393 & 0.2512 & 0.1428 & 0.1036 \\
    MKR & 0.1447 & 0.0850 & 0.1073 & 0.2245 & 0.0863 & 0.0575 \\
    KTUP & 0.1891 & 0.1566 & 0.1579 & 0.3230 & 0.1761 & 0.1299 \\\vspace{3pt}
    ConvE-Rec & 0.2426 & 0.1742 & 0.1993 & 0.3676 & 0.1850 & 0.1357 \\
    \name & 0.2201 & 0.1552 & 0.1889 & 0.3543 & 0.1716 & 0.1266\\
    \name* & \textbf{0.2483} & \textbf{0.1766} & \textbf{0.1994} & \textbf{0.3699} & \textbf{0.1874} & \textbf{0.1371}\\
    \midrule
    Gain over KTUP & 31.31\% & 13.41\% & 26.28\% & 14.52\% & 6.41\% & 5.54\%\\  
    \bottomrule 
    \end{tabular}
}
\centering\caption{Performance of different models on three datasets.
The gain of \name* over KTUP is statistically significant at the 0.001 level according to t-test.
}\label{tab::comparision}
\end{table*}

In this section, we evaluate \name\ on three real-world datasets\footnote{We will make the code and processed datasets public if our paper gets accepted.}. Compared to other state-of-the-art methods, our proposed approach has the following advantages:
\begin{itemize}[leftmargin=*]
    \item \textbf{Effectiveness}. \name\ significantly outperforms existing state-of-the-art KG-based recommendation methods in terms of recommendation accuracy.
    \item \textbf{Explainability}. Case studies on generated paths demonstrate that \name\ can offer good explanations for recommended items.
\end{itemize}


Next, we first describe the evaluation datasets and experimental set-ups.

\subsection{Data and Experiment Settings}

\subsubsection{Data}
We test \name\ on three benchmark datasets for KG-based recommendation: 
\begin{itemize}[leftmargin=*]
\item \textit{Last.FM}\footnote{https://grouplens.org/datasets/hetrec-2011/}. 
This dataset contains a set of music artist listening information from a popular online music system Last.Fm. 

\item \textit{MovieLens-1M}\footnote{https://grouplens.org/datasets/movielens/1m/}. 
MovieLens-1M provides users' ratings towards thousands of movies. For these two datasets, we convert the explicit ratings into implicit feedback where each observed rating is treated as ``1'', and unobserved ratings are marked as ``0''s. Following \cite{Wang_2019}, we use Microsoft Satori to construct knowledge graphs for Last.FM and MovieLens-1M datasets respectively.
\item \textit{DBbook2014}\footnote{http://2014.eswc-conferences.org/important-dates/call-RecSys.html}. 
This dataset provides users' reading history in the book domain. Its supporting knowledge graph is extracted from DBpedia.
\end{itemize}

As we focus on KG-based recommendation, we remove items that have no matching entities in the corresponding knowledge graph. The statistics of processed datasets are presented in Table~\ref{tab::dataset}. Following \cite{Cao_2019,wang2018ripplenet,Wang2018ExplainableRO}, we randomly split the interactions of each user into training, validation, and test set with ratio 6:2:2.


\subsubsection{Baseline Methods} 
We compare \name\ with two kinds of methods: 
1) Classical similarity-based methods including:  
\textbf{ItemKNN}, which recommends items that are most similar to target user's historical items; 
\textbf{BPR-MF}~\cite{yu2014personalized}, which is a widely-used matrix factorization method using Bayesian Personalized Ranking (BPR) loss. 
2) KG-based recommendation methods including: 
\textbf{RippleNet}~\cite{wang2018ripplenet}, which propagates users' interests over knowledge graph with attention mechanism;  
\textbf{CFKG}~\cite{zhang2018learning}, which learns users' and items' representations by applying TransE~\cite{bordes2013translating} on the graph $\mathcal{G}'$; 
\textbf{MKR}~\cite{Wang_2019}, which learns both user-item matching task and knowledge graph embedding task under multi-task learning framework; 
\textbf{KTUP}~\cite{Cao_2019}, which is a state-of-the-art KG-based recommender that jointly learns translation-based recommendation~\cite{he2017translation} and translation-based knowledge graph embedding;
\textbf{ConvE-Rec}, which learns users' and items' embeddings based on integrated graph $\mathcal{G}'$ with ConvE~\cite{dettmers2018convolutional}. As we use ConvE model for reward shaping, we treat ConvE-Rec as a special recommender.  

\subsubsection{Evaluation Metrics} Following~\cite{Cao_2019,he2015trirank,song2019session}, we adopt \textit{Hit Ratio} (HR) and \textit{Normalized Discounted Cumulative Gain} (NDCG) to evaluate the effectiveness of proposed \name\ and baseline methods. We use the same definition of HR and NDCG in \cite{he2015trirank}, where HR measures whether test items are present in the recommendation list and NDCG assesses the ranking quality of test items respectively.
In our study, we always report the averaged HR@K and NDCG@K scores across all users over five runs.

\subsubsection{Implementation Details} First of all, we add $<e_t, r^{-1}, e_h>$ into $\mathcal{G}'$ if a triplet $<e_h, r, e_t>$ exists to enhance the connectivity of graph, where $r^{-1}$ is the inverse of relation $r$.
Following~\cite{Cao_2019}, we only preserve those triplets that are directly connected to items in each supporting knowledge graph.
Since the volume of action space $\mathcal{A}_t$ can be quite large for some nodes, it's time- and memory-consuming to maintain the size of action space based on the largest action space for all states. To simply implementation and computation, we only preserve a fixed number (256 in our experiments) of ``important'' outgoing edges for each node, and the importance of edges are decided by the PageRank scores~\cite{Pageetal98} of their end nodes. For nodes that have less than 256 outgoing edges, we up-sampling their outgoing edges to 256. 

We implement \name\ with Pytorch~\cite{paszke2017automatic}. 
Entity and relation embeddings are pre-trained by applying ConvE~\cite{dettmers2018convolutional}
on graph $\mathcal{G}'$,
and the embedding size is set to 32 for all methods except for ItemKNN, which has no latent representations. Meanwhile, we use the score function of ConvE to compute augmented rewards (i.e., Equation~\ref{eq:reward}). 
From Figure~\ref{fig:example}, we can see path patterns ``User->Item->Entity->item'' and ``User->Item->User->Item'' are more probable to be meaningful, so we empirically set the maximum path length $T$ to 3 as our default setting. We select action dropout rate from \{0.1-0.9\}, dropout rate for entity/relation embeddings from \{0.1-0.9\} using grid search.
Meanwhile, grid search is also applied to select the optimal hyper-parameters for other baseline methods based on the performance on validation data.
For training, we use Adam~\cite{kingma2014adam} optimizer for all models with batch size of 512.
For recommendation, we use beam search with beam size 64 to generate paths for target users. For duplicate paths leading to the same item, we keep the one with the highest probability. 
Finally, we adopt two different ranking strategies to generate final top-K recommendation list: (1) ranks the searched items according to the path probabilities and we denote it as \name, 2) ranks the searched items based on ``rewards'' defined by $\sigma(\psi(e_0, e_T))$ in Equation~\ref{eq:reward} and we denote it as \name*.

\subsection{Analysis of Recommendation Performance}
We report the recommendation accuracy of different methods in Table~\ref{tab::comparision}. We can see that KG-based recommendation methods consistently outperform classical similarity-based methods, which indicates that knowledge graphs indeed help to alleviate the problem of data sparsity in the recommendation. Among KG-based recommendation methods, the RippleNet performs worst, which may be attributed to representing users with multi-hop away entities. KTUP performs strongly because it takes the advantages of both translation-based recommendation and multi-task learning. Note that the only difference between ConvE-Rec and CFKG is the used knowledge graph embedding methods; however, ConvE-Rec achieves much better performance. The reason behind this is that ConvE is a state-of-the-art knowledge graph embedding method, which outperforms TransE used in CFKG. By using pre-trained ConvE embeddings to augment rewards, our \name\ and \name* significantly outperform existing state-of-the-art KG-based recommendation methods in most cases and perform comparably to the unexplainable ConvE-Rec model, which shows that our proposed methods are quite effective.  

\begin{table*}[tb]
\centering
\scalebox{0.93}
{
\begin{tabular}{lcccccc}
\toprule
\multirow{2}{*}{Model} & \multicolumn{2}{c}{Last.FM} & \multicolumn{2}{c}{MovieLens-1M} & \multicolumn{2}{c}{DBbook2014} \\
& HR@10 & NDCG@10 & HR@10 & NDCG@10 & HR@10 & NDCG@10 \\
\midrule 
\vspace{3pt}
\name\ & \textbf{0.2201} & \textbf{0.1552} & \textbf{0.1889} & \textbf{0.3543} & \textbf{0.1716} & \textbf{0.1266} \\
\name-KG & 0.2061 & 0.1466 & 0.1869 & 0.3489 & 0.0802 & 0.0525 \\
\name-RS & 0.0614 & 0.0349 & 0.0654 & 0.1132 & 0.1174 & 0.0867 \\
\name-AD & 0.1350 & 0.0827 & 0.1715 & 0.3217 & 0.1449 & 0.1083 \\ 
\name(T=5) & 0.2108 & 0.1505 & 0.1859 & 0.3500 & 0.1524 & 0.1125 \\
\bottomrule
\end{tabular}
}
\centering\caption{Performance w.r.t. model variants, where [-] means removing that component from the \name.}\label{tab::ablation}

\end{table*}

\begin{figure}[t]
    \centering
    \includegraphics[width=0.95\linewidth]{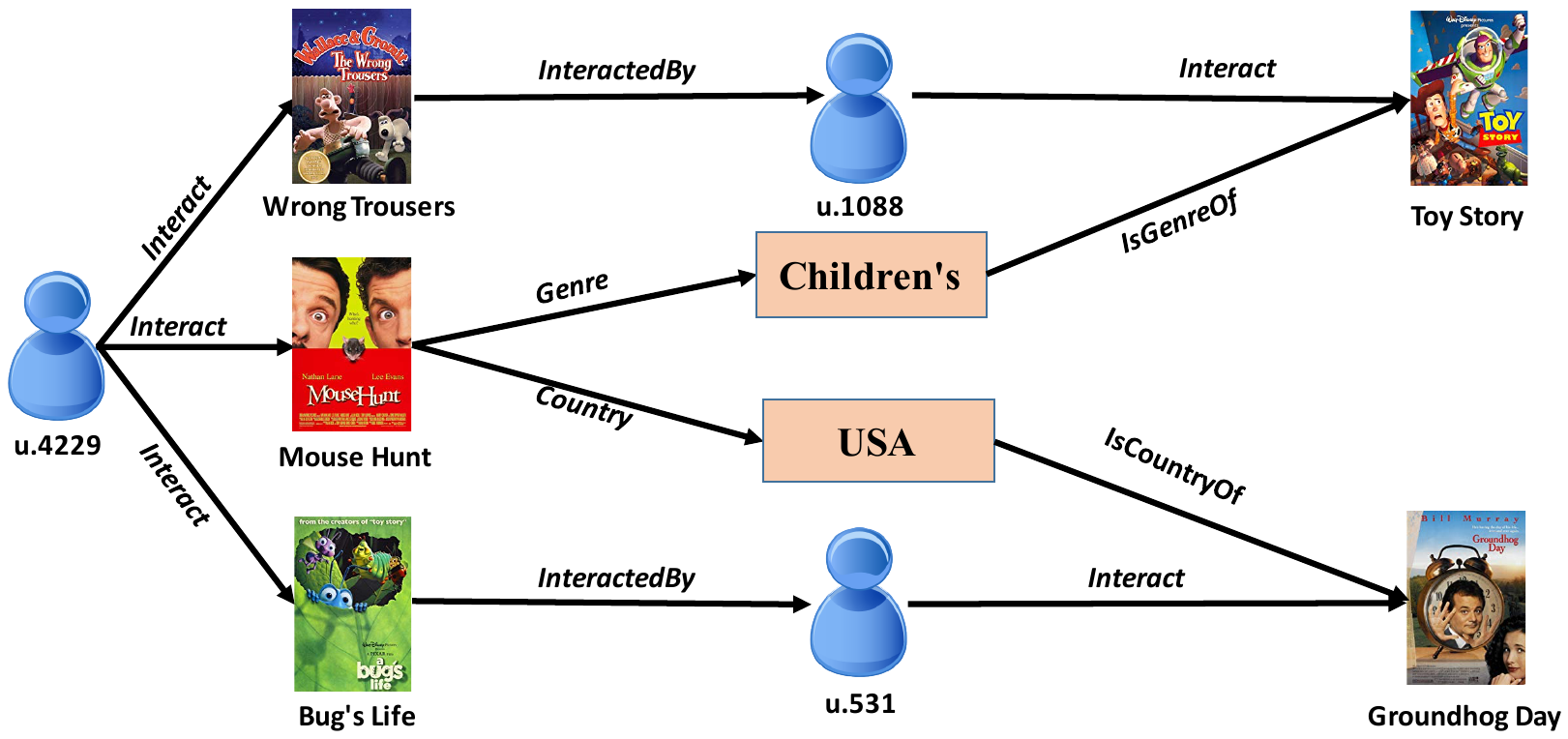}
    \caption{Recommendations and explanations for user \#4229 in MovieLens-1M dataset. ``Toy Story'' is recommended because 1) it shares the same movie genre(i.e., Children's) with ``Mouse Hount``, and 2) it is watched by another user who also watched ``Wrong Trousers''. These explainable paths are automatically discovered by \name\ in a generative manner. 
    }
    \label{fig::case}
\end{figure}

\subsection{Analysis of Explainability}
After demonstrating the effectiveness of \name, we now illustrate its explainability, which is the main contribution of this work to recommender systems. As introduced in previous sections, \name\ provides recommendations by generating meaningful paths from users to items, where paths serve as explanations for recommended items. To give you an intuitive example, we randomly select a real user from MovieLens-1M dataset and search preference paths for them with \name.
As shown in Figure~\ref{fig::case}, we can easily understand that ``Toy Story'' is recommended because it shares the same genre (i.e., Children's) with ``Mouse Hunt'', which the user watched before.
What'more, we find that \name\ model tends to provide multiple explanations for the recommendations. For example, movie ``Toy Story'' is also liked by another user (\#1088) who watched ``Wrong Trousers''. Such diverse explanations may better motivate users to accept the corresponding recommendations.

Beyond explanations for an individual user, we are also interested in global preference path patterns discovered by \name. More specifically, we try to figure out what are the typical path patterns w.r.t. different datasets. 
 From Table~\ref{tab::case}, we can see that \name\ relies more on the path pattern ``User$\xrightarrow{Interact}$Movie$\xrightarrow{Interact^{-1}}$User$\xrightarrow{Interact}$Movie'' on MovieLens-1M dataset.
 Interestingly, we find that \name\ learns relatively diverse path patterns on DBbook2014 dataset such as ``User$\xrightarrow{Interact}$Book$\xrightarrow{LinkedTo}$WikiPage$\xrightarrow{Link}$Book'' and \\
 ``User$\xrightarrow{Interact}$Book$\xrightarrow{Type}$Type$\xrightarrow{IsTypeOf}$book'', which lead to new books that share the same WikiPage or Type with users' historical books respectively. The reason behind the discrepancy of path patterns on two datasets may be two folds. First, note that the average number of interactions for each user in MovieLens-1M data is about 100 while this number is 12 in DBbook2014 data. Therefore it is easier to find a user sharing similar movie preference in MovieLens-1M data than to find a user with similar reading taste in DBbook2014 data. 
Second, the size of supporting knowledge graph for DBbook2014 data is much larger than the number of user-item interactions, so our \name\ learns to make more use of external knowledge when the user-item interactions are sparse.

\begin{table}[tb]
    \centering
    \begin{tabular}{cc}
  \toprule
  Explainable preference paths & Pct. (\%)  \\ 
  \midrule
  \text{U}$\xrightarrow{Int.}$\text{M}$\xrightarrow{Int.^{-1}}$\text{U}$\xrightarrow{Int.}$\text{M} & 78.6 \\
  \text{U}$\xrightarrow{Int.}$\text{M}$\xrightarrow{Country}$\text{C}$\xrightarrow{IsCountryOf}$\text{M} & 15.4 \\
  \text{U}$\xrightarrow{Int.}$\text{M}$\xrightarrow{Genre}$\text{G}$\xrightarrow{IsGenreOf}$\text{M} & 2.3 \\
   \midrule
   \text{U}$\xrightarrow{Int.}$\text{B}$\xrightarrow{Type}$\text{T}$\xrightarrow{IsTypeOf}$\text{B} & 64.9 \\
   \text{U}$\xrightarrow{Int.}$\text{B}$\xrightarrow{LinkedTo}$\text{WP}$\xrightarrow{Link}$\text{B} & 21.7 \\
  \text{U}$\xrightarrow{Int.}$\text{B}$\xrightarrow{Int.^{-1}}$\text{U}$\xrightarrow{Int.}$\text{B} & 7.4 \\

   \bottomrule
  \end{tabular}
    \caption{Most frequent path patterns of Ekar during inference on MovieLens-1M (top) and DBbook2014 (bottom) datasets. ``U'', ``M'', ``C'', ``G'', ``B'', ``T'' and ``WP'' represent User, Movie, Country, Genre, Book, Type and WikiPage respectively. ``$Int.$'' and ``$Int.^{-1}$'' are ``Interact'' and  ``Interacted$^{-1}$'' in short respectively.}
    \label{tab::case}
\end{table}

\subsection{Ablation Study}\label{subsec:ablation}
In this section, we compare different variants of \name\ to show the influences of some essential components such as KG, reward shaping, action dropout and maximum path length $T$, which are denoted as \name-KG, \name-RS, \name-AD and \name(T=5) respectively in Table~\ref{tab::ablation}. 
We find the influence of KG is very significant on Last.FM and DBbook2014 datasets. This is because these two datasets are extremely sparse, while the MovieLens-1M data is relatively dense. 
Removing reward shaping leads to a severe performance drop on all datasets because \name\ without reward shaping assigns zero rewards to all items that a user has not interacted with. In this way, the agent is penalized for exploring potential items that are of interest to users and therefore cannot effectively generate recommendations.
Besides reward shaping, we also use action dropout to further encourage the agent to explore diverse paths, and we can see that \name-AD performs worse than the full model \name. 
At last, we try a larger maximum path length $T$ to enable the agent to explore longer paths. We find that \name(T=5) performs worse than \name\ on all datasets.
This is because long paths may introduce more noise and thus be less meaningful. However, thanks to the stop mechanism, the performance drop is not significant. 

To better understand the role of stop action, we present the details about \name(T=5)'s paths patterns in Table~\ref{tab::T5_case}. We have two observations. First, there are many paths of length five, which means that our agent is recommending items that have high-order similarity to users' historical items. 
Second, since the supporting knowledge graph is very large on DBbook2014 dataset and paths with length five may not be always useful, our agent learns to infer path with length of three or four by taking stop actions.


\begin{figure*}[th]
\includegraphics[width=0.98\textwidth]{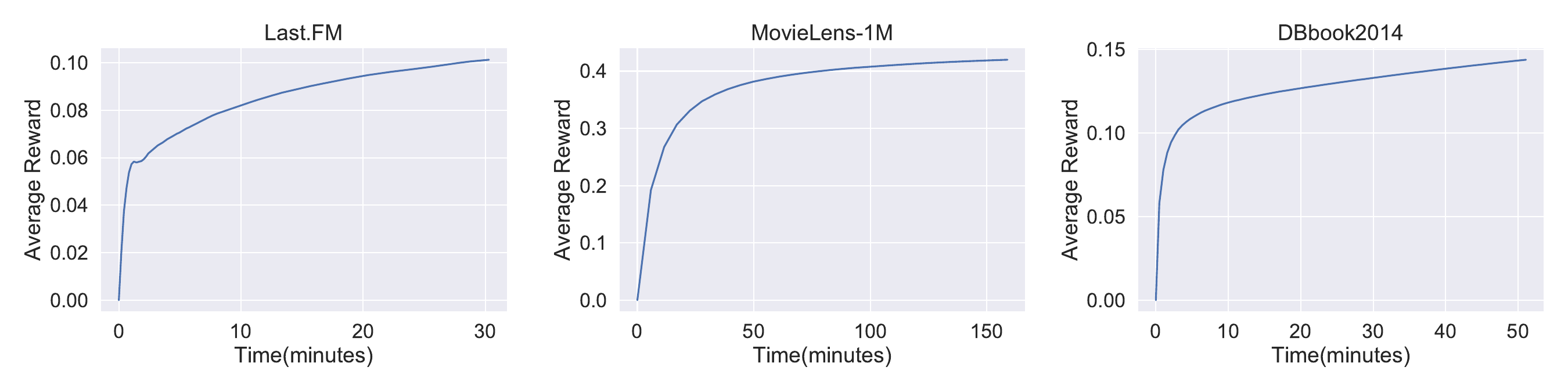}
\caption{Results of running time with batch size of 512 and maximum path length of 3. The x-axis is the training time in minutes, and the y-axis is the average rewards over training samples.
}\label{fig::time_reward}
\end{figure*}

\begin{table}[tb]
\centering
  \begin{tabular}{cc}
  \toprule
  Explainable preference paths & Pct. (\%)  \\ 
  \midrule
  \text{U}$\xrightarrow{Int.}$\text{M}$\xrightarrow{Int.^{-1}}$\text{U}$\xrightarrow{Int.}$\text{M}$\xrightarrow{Int.^{-1}}$U$\xrightarrow{Int.}$M & 58.7 \\
  \text{U}$\xrightarrow{Int.}$\text{M}$\xrightarrow{Int.^{-1}}$\text{U}$\xrightarrow{Int.}$\text{M}$\xrightarrow{Country}$\text{C}$\xrightarrow{IsCountryOf}$\text{M} & 25.7 \\  
  \midrule
  \text{U}$\xrightarrow{Int.}$\text{B}$\xrightarrow{Int.^{-1}}$\text{U}$\xrightarrow{Int.}$\text{B}$\xrightarrow{Type}$T$\xrightarrow{IsTypeOf}$B & 52.3 \\
  U$\xrightarrow{\mathbf{Stop}}$U$\xrightarrow{Int.}$B$\xrightarrow{Type}$T$\xrightarrow{IsTypeOf}$B$\xrightarrow{\mathbf{Stop}}$B & 13.2 \\
  U$\xrightarrow{\mathbf{Stop}}$U$\xrightarrow{Int.}$B$\xrightarrow{Type}$T$\xrightarrow{IsTypeOf}$B$\xrightarrow{Link}$B  & 9.9 \\
  \bottomrule
  \end{tabular}
  \caption{Most frequent path patterns of \name(T=5) during inference on MovieLens-1M (top) and Last.FM (bottom) datasets. 
``U'', ``M'', ``C'', ``B'' and ``T'' represent User, Movie, Country, Book and Type respectively. ``$Int.$'' and ``$Int.^{-1}$'' are ``Interact'' and  ``Interacted$^{-1}$'' in short respectively.
}\label{tab::T5_case}
\end{table}

\subsection{\name\ with Different KGE Methods}\label{sec::distmult}
Although \name\ model has been utilizing ConvE model so far to pre-train entity and relation embeddings for both initialization and reward shaping, it is notable that \name\ is independent of any specific knowledge graph embedding methods. 
Therefore we also test our model with another widely-used knowledge graph embedding method DistMult~\cite{yang2014embedding}. The score functions of DistMult and ConvE are presented in Table~\ref{tab::score_func}. For a fair comparison, we use the same experimental settings and just substitute DistMult for ConvE in our experiments. Following ConvE-Rec, we denote recommendation with DistMult as DistMult-Rec.

\begin{table}[t]
    \centering
    \begin{tabular}{cc}
    \toprule
        Model & Score function $\psi(e_0, e_T)$  \\
        \midrule
        DistMult & $\langle \mathbf{e}_0, \mathbf{r}, \mathbf{e}_T \rangle$ \\
        ConvE & $g(\text{vec}(g([\overline{\mathbf{e}_0};\overline{\mathbf{r}}]* \omega)) \mathbf{W})\mathbf{e}_T$ \\
    \bottomrule
    \end{tabular}
    \caption{Score functions w.r.t. different KGE methods, where $\langle \cdot \rangle$ denotes generalized inner product of three vectors, $\overline{\cdot}$ denotes a 2D shaping of vectors, $*$ is the convolution operator, $\omega$ denotes filters in convolutional layers, $g(\cdot)$ is a non-linear activation function and $\text{vec($\cdot$)}$ converts a tensor to a vector. $\mathbf{e}_0$, $\mathbf{e}_T$ and $\mathbf{r}$ are embeddings of user $e_0$, entity $e_T$ and relation ``Interact'' respectively.}
    \label{tab::score_func}
\end{table}

The results of using different knowledge graph embedding methods on MovieLens-1M and DBbook2014 are presented in Table~\ref{tab::distmult}. First, we can see that ConvE-Rec outperforms DistMult-Rec on these two datasets. This is because ConvE has proven more effective than DistMult for knowledge graph completion, as a result of which our \name\ with ConvE also outperforms \name\ with DistMult. Second, \name\ (DistMult) and \name* (DistMult) outperform DistMult-Rec on MovieLens-1M dataset and perform comparably to DistMult-Rec on DBbook2014 dataset, which is consistent with the observations of \name\ (ConvE) and \name* (ConvE). We omit the results on Last.FM dataset because they show similar trends and lead to the same conclusion.

\begin{table}[!tb]
\scalebox{0.95}
{
\begin{tabular}{lcccc}
\toprule
\multirow{2}{*}{Model} & \multicolumn{2}{c}{MovieLens-1M} & \multicolumn{2}{c}{DBbook2014} \\
 & HR@10 & NDCG@10 & HR@10 & NDCG@10 \\
\midrule 
\vspace{3pt}
ConvE-Rec &  0.1993 & 0.3676 & 0.1850 & 0.1357 \\
\vspace{3pt}
DistMult-Rec  & 0.1773 & 0.3341 & 0.1535 & 0.1090 \\
\name\ (ConvE)  & 0.1889 & 0.3543 & 0.1716 & 0.1266\\\vspace{3pt}
\name\ (DistMult) & 0.1761 & 0.3352 & 0.1367 & 0.0958 \\
\name* (ConvE) & {0.1994} & {0.3699} & {0.1874} & {0.1371}\\
\name* (DistMult) & 0.1774 & 0.3343 & 0.1482 & 0.1061 \\
\bottomrule
\end{tabular}
}
\centering\caption{Effectiveness comparison of using different knowledge graph methods for entity/relation initialization and reward shaping.
The trends on Last.FM dataset are similar hence omited. 
}
\label{tab::distmult}
\end{table}

\subsection{Convergence Analysis}
We present the running time of \name\ in Figure~\ref{fig::time_reward}. As can be seen, \name\ converges fast on MovieLens-1M dataset with less than ten minutes, while it takes a bit more time to converge on the other two datasets. The reason for different convergence behaviors is that it is easier to walk to correct items on dense datasets (e.g., MovieLens-1M) than on sparse datasets (e.g., DBbook2014). Overall, our \name\ is efficient because we initialize entity/relation embeddings with pre-trained knowledge graph embeddings, and we use reward shaping to augment reward signals.

%% file: 6-conclusion.tex
\section{Conclusion}\label{sec::conclusion}
In this paper, we introduced a novel approach to provide explanations for recommendation with knowledge graphs.
Our proposed \name\ generates meaningful paths from users to relevant items by learning a walk policy on the user-item-entity graph.
Experimental results show that \name\ outperforms existing KG-based recommendation methods and is quite efficient. Furthermore, we demonstrate the explainability of \name\ via insightful case studies on different datasets. Future work includes incorporating domain knowledge to design proper reward functions for recommendation task and developing a distributed version of \name\ for even larger datasets.

\section*{Acknowledgments}
The authors would like to thank Meng Qu and Zafarali Ahmed for providing useful feedback on initial versions of the manuscript. We also thank Yue Dong and Zhaocheng Zhu for editing the manuscript. WS and MZ are partially supported by Beijing Municipal Commission of Science and Technology under Grant No. Z181100008918005 as well as the National Natural Science Foundation of China (NSFC Grant Nos.61772039 and 91646202). WS also acknowledges the financial support by Chinese Scholarship Council. JT is supported by the Natural Sciences and Engineering Research Council of Canada, as well as the Canada CIFAR AI Chair Program.